\documentclass[prb,twocolumn,floatfix,showpacs,amssymb]{revtex4}

\usepackage{graphicx}
\usepackage{dcolumn}
\usepackage{bm}
\begin{document}

\title{Contribution of weak localization to
non local transport \\at normal metal / superconductor
double interfaces}
\author{R. M\'elin}
\affiliation{
Centre de Recherches sur les Tr\`es Basses
Temp\'eratures, CRTBT\cite{crtbt},\\ CNRS, BP 166,
38042 Grenoble Cedex 9, France}

\begin{abstract}
In connection
with a recent experiment [Russo {\it et al.},
Phys. Rev. Lett. {\bf 95},  027002 (2005)],
we investigate the effect of weak localization on 
non local transport in
normal metal / insulator / superconductor / insulator
/ normal metal (NISIN) trilayers,
with extended interfaces.
The negative weak localization contribution to the crossed resistance
can exceed in absolute value
the positive elastic cotunneling contribution
if the normal metal
phase coherence length or the energy are large enough.
\end{abstract}

\maketitle

\section{Introduction}
The manipulation of
correlated pairs of electrons in solid state
devices has aroused a considerable interest recently.
The goals of this line of research
are the realization of sources of entangled pairs of electrons
for quantum information, and the 
realization of fundamental tests
of quantum mechanics
\cite{Choi,Martin}.
Correlated pairs of electrons
can be extracted from a superconductor by
Andreev reflection, 
with extended or localized interfaces
between superconductors and normal metals or ferromagnets
\cite{Byers,Deutscher,Samuelson,Prada,Koltai,Beckmann,Russo},
and in Josephson junctions involving a double
bridge between two superconductors \cite{Melin-Peysson}.

Charge is 
transported by Andreev reflection at a normal metal~/
superconductor (NS) interface:
an electron coming from
the normal side is reflected as a 
hole while a Cooper pair is transmitted
in the superconductor \cite{Andreev}. 
In the superconductor,
Andreev reflection is mediated by an evanescent state of linear
dimension set by the superconducting
coherence length~$\xi$. In structures in which 
a superconductor is multiply
connected to normal metal electrodes separated by
a distance of order $\xi$
 \cite{Lambert,Jedema}, the Andreev reflected hole
in the spin-($-\sigma$) band
can be transmitted in an electrode different from the 
one in which the incoming spin-$\sigma$ electron propagates.
This ``non local'' transmission in the electron-hole channel
is called ``crossed Andreev reflection''. Non
local transmission in the electron-electron channel corresponds to
``elastic cotunneling'' by which an electron is
transmitted from one electrode to another while spin is conserved \cite{Falci}.

\begin{figure}
\includegraphics [width=1. \linewidth]{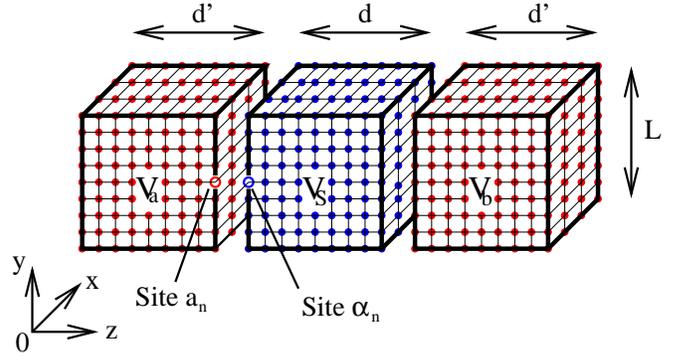}
\caption{(Color online.) Schematic representation of the tight-binding
model of NISIN trilayer. The insulating layers are not shown on the figure.
The aspect ratio is not to the scale of the experiment
in Ref.~\onlinecite{Russo} for which $d=15,\,50,\,200\,$nm,
$L$ is of order of $5\,\mu$m, and $d'\simeq 50\,$nm.
\label{fig:trilayer}
}
\end{figure}

A schematic three-terminal device is represented on Fig.~\ref{fig:trilayer},
as well as the voltages $V_a$ and $V_b$ 
applied on the normal electrodes ``a'' and ``b'' respectively.
The voltage $V_S$ on the superconductor is chosen
as the reference voltage ($V_S=0$).
Non local transport is characterized by a current $I_a(V_b)$
circulating in electrode ``a'' in
response to a voltage $V_b$ on electrode ``b''.
It is supposed 
in addition that $V_a=0$: electrode ``a'' is grounded,
like in experiments \cite{Beckmann,Russo}
(see Fig.~\ref{fig:trilayer}).
The crossed conductance  is defined by
${\cal G}_{a,b}(V_b) =
\partial I_{a}/\partial V_{b}(V_b)$.
A crossed conductance dominated by elastic cotunneling
(crossed Andreev reflection)
is negative (positive) \cite{note-R}
because of the opposite charges of the outgoing particle in 
elastic cotunneling and crossed Andreev reflection.
Lowest
order perturbation theory in the tunnel amplitudes leads to
${\cal G}_{a,b}(V_b) =0$ because
the crossed Andreev reflection and elastic cotunneling 
channels have in this case
an exactly opposite contribution to the crossed
conductance once the average over
disorder \cite{Feinberg-des,Cht}
or over
the Fermi oscillations in multichannel
ballistic systems \cite{Falci,EPJB} is taken into account.

Three unexpected experimental
features for the crossed conductance
in a normal metal~/ insulator~/ superconductor~/ insulator~/
normal metal (NISIN)
trilayer have been reported recently
by Russo {\sl et al.} \cite{Russo}:
i) The crossed conductance does not average to zero with normal
metals, in contradiction to the abovementioned
prediction of lowest order perturbation
theory in the tunnel amplitudes \cite{Falci}. The
order of magnitude of the
experimentally observed crossed signal is 
far from being compatible with
lowest order perturbation theory.
ii) A
magnetic field applied parallel to the interfaces suppresses
the non local signal,
suggesting a phase coherent process.
iii) The sign of the crossed resistance \cite{note-R} 
crosses over from positive (the sign of elastic cotunneling) 
to negative (the sign of crossed Andreev reflection)
as the bias voltage $V_b$ increases, and
the crossed signal disappears
if the bias voltage exceeds the Thouless energy in the
superconductor.

We show here that 
weak localization with extended interfaces leads to a positive
crossed conductance, the sign of which is opposite to the sign
of the dominant elastic cotunneling channel for
localized interfaces \cite{Melin-Feinberg}. The weak localization
contribution to the crossed conductance becomes important at
large bias voltages and for a large phase coherence
length in the normal metals.

The article is organized as follows. Necessary preliminaries
are given in Sec.~\ref{sec:prelim}.
The weak localization contribution to crossed transport
is discussed in Sec.~\ref{sec:nonlocal}.
Concluding remarks are given in Sec.~\ref{sec:conclu}.

\section{Preliminaries}
\label{sec:prelim}
\subsection{Hamiltonians}
\label{sec:ham}
The normal metal electrodes are described by the tight binding
Hamiltonian
\begin{equation}
\label{eq:HN}
{\cal H}_{\rm N}=\sum_{{\bf k},\sigma}
\epsilon({\bf k}) c_{{\bf k},\sigma}^+ c_{{\bf k},\sigma}
,
\end{equation}
where ${\bf k}$ is the wave-vector,
$\sigma=\uparrow,\downarrow$ the
projection of the electron spin on the spin quantization axis,
and where
$\epsilon({\bf k})=-2 t_0 [\cos{(k_x a_0)}+\cos{(k_y a_0)}
+\cos{(k_z a_0)}]$ is the 
dispersion relation of free electrons on a cubic lattice,
with $t_0$ the bulk hoping amplitude,
$a_0$ the distance between two neighboring sites, and
$k_x$, $k_y$ and $k_z$ the projections of the electron wave-vector
on the $x$, $y$ and $z$ axis (see Fig.~\ref{fig:trilayer}).

The superconductor is described by the BCS Hamiltonian
\begin{equation}
\label{eq:H-BCS}
{\cal H}_{\rm BCS}=\sum_{{\bf k},\sigma}
\epsilon({\bf k}) c_{{\bf k},\sigma}^+ c_{{\bf k},\sigma}\\\nonumber
+ \Delta \sum_{\bf k} \left(
c_{{\bf k},\uparrow}^+ c_{-{\bf k},\downarrow}^+
+ c_{-{\bf k},\downarrow} c_{{\bf k},\uparrow}
\right)
,
\end{equation}
with $\Delta$ the superconducting gap.

Diagonal disorder is included \cite{Abrikosov},
with an elastic mean free path $l_e^{(N)}$ in the normal
electrodes of the NISIN trilayer, and $l_e^{(S)}$ in
the superconducting electrode.
A finite coherence length $l_\varphi^{(N)}$ 
in the normal electrodes
is accounted for by adding
phenomenologically an imaginary part to the energy.

At the extended interface $a$-$\alpha$,
the tunnel Hamiltonian connecting the electrodes ``a'' and ``$\alpha$''
takes the form
\begin{equation}
\label{eq:W}
\hat{\cal W}_{a,\alpha}=\sum_{n,\sigma} -t \left(
c_{a_n,\sigma}^+ c_{\alpha_n,\sigma} +
c_{\alpha_n,\sigma}^+ c_{a_n,\sigma} \right)
,
\end{equation}
where the sites $a_n$ on the normal metal side of the interface
correspond to the sites $\alpha_n$ on the superconducting side of the
interface (see Fig.~\ref{fig:trilayer}), and a similar expression
holds for the tunnel Hamiltonian at the interface $b$-$\beta$.
\begin{figure*}
\includegraphics [width=1. \linewidth]{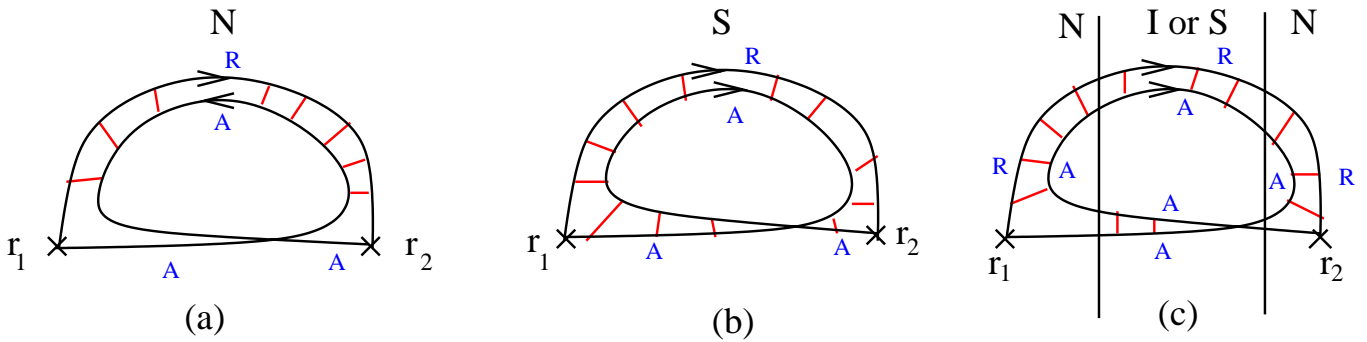}
\caption{(Color online.) A weak localization diagram in a normal metal
for which ${\bf r}_1$ and ${\bf r}_2$ are within
the elastic mean free path $l_e$ (a), and
in a superconductor for which ${\bf r}_1$ and ${\bf r}_2$ are
within $\xi$. (c) represents the same diagram at a NISIN double interface,
equivalent to Fig.~\ref{fig:A-bis}c. In 
the NISIN case on (c) ${\bf r}_1$ and
${\bf r}_2$ are within $\xi$ and ${\bf r}_1$,
${\bf r}_2$ are within $l_e^{(N)}$ from the interfaces.
In the NIN case on (c), ${\bf r}_1$ and ${\bf r}_2$
are within the insulator correlation length.
``A'' and ``R'' stand for advanced and retarded.
\label{fig:d-s}
}
\end{figure*}

\subsection{Green's functions}
\label{sec:Green}
The fully dressed advanced and retarded
equilibrium Nambu Green's function
$\hat{G}^{A,R}(\omega)$ at energy $\omega$ is first determined
by solving the Dyson equation
\begin{equation}
\hat{G}^{A,R}(\omega)=\hat{g}^{A,R}(\omega)+
\hat{g}^{A,R}(\omega)\otimes \hat{\Sigma}_t \otimes
\hat{G}^{A,R}(\omega)
,
\end{equation}
where
$\otimes$ denotes a summation
over the spatial indices, and
$\hat{g}^{A,R}(\omega)$ are the advanced and retarded
Green's functions of the disconnected system ({\it i. e.}
with $t=0$ in the tunnel Hamiltonian given by Eq.~(\ref{eq:W})).
The self-energy $\hat{\Sigma}_t$ corresponds to the couplings
in the tunnel
Hamiltonian given by Eq.~(\ref{eq:W}).
The Green's functions are
$4\times 4$ matrices in the spin $\otimes$ Nambu representation.
The four components correspond to a spin-up electron, a spin-down hole,
a spin-down electron and a spin-up hole. Because of spin rotation
invariance, some elements of the $4\times 4$ Green's functions
are redundant. We work here in a $2 \times 2$
block in the sector $S_z=\hbar/2$,
encoding the superconducting correlations
among a spin-up electron (Nambu label ``1'')
and a spin-down hole (Nambu label ``2'').

Once the fully dressed advanced and retarded Green's functions have
been obtained, the
Keldysh Green's function $\hat{G}^{+,-}$
is determined by the
Dyson-Keldysh
equation\cite{Nozieres,Martin-Rodero93,Levy95-PRB,Levy95-JPCM,Cuevas}
\begin{equation}
\label{eq:Dyson}
\hat{G}^{+,-}=\left(\hat{I}+
\hat{G}^R \otimes \hat{\Sigma}_t \right)
\otimes \hat{g}^{+,-} \otimes
\left(\hat{I}+\hat{\Sigma}_t \otimes \hat{G}^A\right)
,
\end{equation}
where $\hat{g}^{+,-}$ is the
Keldysh Green's function of the isolated electrodes,
and where the energy dependence of the
Green's functions is omitted.

\subsection{Transport formula}

The current through the interface $a$-$\alpha$ is given by
\begin{eqnarray}
\label{eq:I-up}
I_{a,\alpha}&=&\frac{2 e}{h} \int d \omega
\sum_n \mbox{Tr}
\left\{ \left[\hat{t}_{a_n,\alpha_n}
\hat{G}^{+,-}_{\alpha_n,a_n}(\omega)\right.\right.\\
&-& \left.\left.
\hat{t}_{\alpha_n,a_n} \hat{G}^{+,-}_{a_n,\alpha_n}(\omega) \right]
\hat{\sigma}_3 \right\}
,
\nonumber
\label{eq:I-a-al}
\end{eqnarray}
where the trace is a summation over the two components
of the Nambu Green's function,
$\hat{\sigma}_3$ is one of the Pauli matrices, the
diagonal elements of which are $(1,-1)$, and the sum over
$n$ runs over all sites at the interface $a$-$\alpha$. 
As shown on Fig.~\ref{fig:trilayer},
the symbols $a_n$ and $\alpha_n$ in Eq.~(\ref{eq:I-up})
label two corresponding sites at the interface,
in the normal electrode ``a'' and in the superconductor
respectively.
The two spin
orientations are taken into account in the prefactor of
Eq.~(\ref{eq:I-up}).

The local conductance of a NIN interface is equal to
$(e^2/h)T$ per channel, where $T=2\pi^2 t^2
\rho_N^2$ is the dimensionless transmission coefficient
in the tunnel limit,
with $\rho_N$ the normal density of states \cite{Nozieres}.

The non perturbative
transport formula for the local current at a localized NIN
interface was obtained in Ref.~\onlinecite{Nozieres},
and generalized in Ref.~\onlinecite{Cuevas} to a
localized NIS interface, and in Ref.~\onlinecite{Melin-Feinberg}
to non local transport at a double ferromagnet / superconductor
interface.
We deduce from these references the
exact expressions of the local conductance
${\cal G}_{\rm NIN}(V)$ of a single NIN interface,
of the Andreev conductance
${\cal G}_{\rm NIS}(V)$ of a single NIS interface,
and of the
crossed conductance ${\cal G}_{a,b}(V_b)$ of a NISIN trilayer
with extended interfaces:
\begin{eqnarray}
\label{eq:Gloc-NIN}
&&{\cal G}_{\rm NIN}(V) = 8 \pi^2 \frac{e^2}{h} 
t^4 \\
\nonumber
&& \times \sum_{j,k,l}
\overline{
\rho^N_{a_i,a_j} 
G_{\alpha_j,\beta_k}^{A,1,1}(e V)
\rho^N_{b_k,b_l}
G_{\beta_l,\alpha_i}^{R,1,1}(e V)}\\
\label{eq:Gloc-def}
&&{\cal G}_{\rm NIS}(V) = 16 \pi^2 \frac{e^2}{h} 
t^4 \\
\nonumber
&& \times \sum_{j,k,l}
\overline{
\rho^N_{a_i,a_j} 
G_{\alpha_j,\beta_k}^{A,1,2}(e V)
\rho^N_{b_k,b_l}
G_{\beta_l,\alpha_i}^{R,2,1}(e V)}\\
\label{eq:Gab-def}
&&{\cal G}_{a,b}(V_b) = -8 \pi^2 \frac{e^2}{h} t^4 \\\nonumber 
&&\times \sum_{j,k,l}
\mbox{Tr}
\left\{
\overline{
\hat{\rho}^N_{a_i,a_j} \hat{\sigma}_3
\hat{G}_{\alpha_j,\beta_k}^{A}(e V_b)
\hat{\rho}^N_{b_k,b_l}
\hat{\sigma}_3
\hat{G}_{\beta_l,\alpha_i}^{R}(e V_b)}
\right\}
,
\end{eqnarray}
valid to all orders in the hopping amplitude.
The crossed conductance
${\cal G}_{a,b}(V_b)$ is per conduction channel
through the junction of area $L\times L$, with a superconducting
layer of thickness $d$ (see Fig.~\ref{fig:trilayer}).
The summations in Eqs.~(\ref{eq:Gloc-def}) and (\ref{eq:Gab-def})
run over all pairs of sites at the interfaces,
and the overline denotes disorder averaging.
The density of states
$\hat{\rho}^N_{a_i,a_j}=(\hat{g}^A_{a_i,a_j}
-\hat{g}^R_{a_i,a_j})/2i\pi$ connects the
sites $a_i$ and $a_j$ in electrode ``a'', and
a similar definition holds for $\hat{\rho}^N_{b_i,b_j}$.
The site $a_j$ ($b_l$) in electrode ``a'' (``b'')
is connected  by the tunnel Hamiltonian to
the site $\alpha_j$ ($\beta_l$) in the superconductor
(see Fig.~\ref{fig:A-bis} for the NISIN interface).
In the case of the local conductance of a NIN junction,
$\alpha_j$, $\beta_k$, $\beta_l$
and $\alpha_i$ belong to an insulating layer that
has been inserted in between the two normal metals.
The tunnel amplitude $t$ in Eq.~(\ref{eq:Gloc-NIN})
connects in this case the normal metal to the insulator, while $t$
in Eqs.~(\ref{eq:Gloc-def}) and (\ref{eq:Gab-def})
connects the normal metal to the superconductor.

The fully dressed
advanced and retarded
equilibrium Nambu Green's functions
$\hat{G}_{\alpha_j,\beta_k}^{A}(\omega)$
and $\hat{G}_{\beta_l,\alpha_i}^{R}(\omega)$
at energy $\omega$ in
Eqs.~(\ref{eq:Gloc-NIN}), (\ref{eq:Gloc-def})
and (\ref{eq:Gab-def})
are expanded by using the Dyson equation
given by Eq.~(\ref{eq:Dyson}).
We deduce from the exact Keldysh transport formula
given by Eqs.~(\ref{eq:Gloc-NIN}), (\ref{eq:Gloc-def})
and (\ref{eq:Gab-def}) that the diagrams are connected,
and that the propagators for the density of states
are directly connected
to at least one tunneling vertex.
The other extremity of the density of states 
propagators is connected either
to a tunneling vertex, or to a disorder scattering vertex.
The density of states, represented by wavy lines on the
diagrams, connects to one advanced and one retarded Green's function
as in Eqs.~(\ref{eq:Gloc-NIN}), (\ref{eq:Gloc-def})
and (\ref{eq:Gab-def}).

A crossed conductance dominated by elastic cotunneling
(crossed Andreev reflection)
is negative (positive) \cite{note-R}.
This can be seen most simply by making the Nambu labels
explicit in Eq.~(\ref{eq:Gab-def}) and taking into account the
signs in the $\hat{\sigma}_3$ matrices, and the global
sign.

The crossed conductance is expanded perturbatively in
$T^2$, where
the normal local transmission coefficient $T$ has already
been defined, and is also expanded
in the number of non local processes crossing the superconductor
since $d \agt \xi$ \cite{Falci}. 

\subsection{Weak localization in a superconductor}
\label{sec:wl}
Weak localization in a superconductor was already investigated
in Ref.~\onlinecite{Smith} in connection with 
the determination of the number density of
superconducting electrons. 
A weak localization diagram in a bulk normal metal is shown on
Fig.~\ref{fig:d-s}a. In this case the two points
${\bf r}_1$ and ${\bf r}_2$ are within the elastic
mean free path $l_e^{(N)}$.
In a bulk superconductor (see Fig.~\ref{fig:d-s}b),
the two points ${\bf r}_1$ and
${\bf r}_2$ are within the coherence length $\xi$
since the disorder average $\overline{g_{\alpha,\beta}^A
g_{\alpha,\beta}^A}$ of two advanced Green's functions
between the two sites $\alpha$
and $\beta$ at ${\bf r}_1$ and ${\bf r}_2$
in a superconductor
is limited by $\xi$, not by $l_e$.
Similar diagrams for the conductance
are introduced at a NISIN double interface in
Sec.~\ref{sec:nonlocal}. In this case the points ${\bf r}_1$ and
${\bf r}_2$ are separated by the superconductor thickness,
of order $\xi$ (see Fig.~\ref{fig:d-s}c).
At a NIN interface, the two
points ${\bf r}_1$ and ${\bf r}_2$ on different interfaces
can be separated by a
distance equal to the thickness of the insulator,
comparable to the decay length induced by the
charge gap of the insulator.

\section{Non local transport in a NISIN trilayer}
\label{sec:nonlocal}

\subsection{Crossed conductance to order $T^2$}
\label{sec:T2}
\begin{figure}
\includegraphics [width=.9 \linewidth]{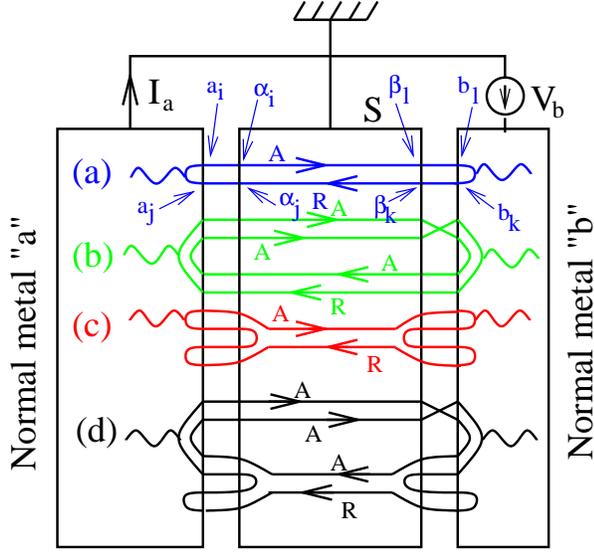}
\caption{(Color online.) The lowest
order diagrams contributing to crossed transport.
We show (a) the vanishing
diagram of order $T^2$, (b) the vanishing weak localization
diagram of order
$T^4$, (c) the first non vanishing diagram
of order $T^4$, and (d) the weak localization diagram of order $T^6$.
The wavy lines correspond to the insertion of the
normal metal density of states,
and of a $\hat{\sigma}_3$ matrix.
\label{fig:A-bis}
}
\end{figure}

Now we evaluate the lowest order diagrams appearing in 
the crossed conductance of a NISIN trilayer.
The crossed conductance due to the diagram of order $T^2$
(see Fig.~\ref{fig:A-bis}a)
vanishes because the contributions
of the elastic cotunneling and
crossed Andreev reflection channels are exactly opposite
\cite{Feinberg-des,Melin-Feinberg,Falci,EPJB}.
This can be seen also by evaluating the summation over
the Nambu labels in the diagram on Fig.~\ref{fig:A-bis}
and using $\overline{g_{\alpha,\beta}^{1,1}g_{\beta,\alpha}^{1,1}}
=\overline{g_{\alpha,\beta}^{1,2}g_{\beta,\alpha}^{2,1}}$, where
$\alpha$ and $\beta$ are two points in the superconductor
at a distance of order $\xi$.

\subsection{Crossed conductance to order $T^4$}
The first weak localization diagram  of order $T^4$
involving four Green's functions in the superconductor
is shown on Fig.~\ref{fig:A-bis}b.
The corresponding crossed conductance
vanishes, as for the diagram of order
$T^2$ discussed in Sec.~\ref{sec:T2}.

The diagram of order $T^4$ on Fig.~\ref{fig:A-bis}c 
takes a finite
value, and is evaluated
explicitely by summing over all possible Nambu labels, and over the
different possibilities of inserting the density of states and the
$\hat{\sigma}_3$ matrices (represented by the wavy lines on the
diagrams). This leads to the crossed conductance
\begin{eqnarray}
\label{eq:crossed-T4}
{\cal G}_{a,b}(V_b)&=&4\pi^2 t^8 \rho_N^6 \sum_\beta
\left(\,
\overline{g_{\alpha,\beta}^{1,1,A}(e V_b)
g_{\beta,\alpha}^{2,2,R}(e V_b)}\right.\\
&-& \left.\overline{g_{\alpha,\beta}^{1,2,A}(e V_b)
g_{\beta,\alpha}^{1,2,R}(e V_b)}\,
\right) \frac{\Delta^2}{\Delta^2
-(e V_b)^2}
\nonumber
,
\end{eqnarray}
where\cite{Melin-Feinberg}
\begin{eqnarray}
\nonumber
\overline{g_{\alpha,\beta}^{1,1}(\omega)
g_{\beta,\alpha}^{2,2}(\omega)}
&=&\frac{\pi^2 \rho_S^2}{2} \mu(R)
\exp{\left(-\frac{2 R}{\xi(\omega)}\right)}
\frac{2\omega^2-\Delta^2}{\Delta^2-\omega^2}\\
\nonumber
\overline{g_{\alpha,\beta}^{1,2}(\omega)
g_{\beta,\alpha}^{1,2}(\omega)}
&=&
\frac{\pi^2 \rho_S^2}{2} \mu(R)
\exp{\left(-\frac{2 R}{\xi(\omega)}\right)}
\frac{\Delta^2}{\Delta^2-\omega^2}
,
\end{eqnarray}
with $R$ the distance between the sites $\alpha$ and $\beta$,
and with $\mu(R)=(k_F R)^2$ in the ballistic limit, and
$\mu(R)=k_F^2 R l_e^{(S)}$ in the diffusive limit.
The resulting term of order $T^4$ in the crossed conductance
is given by
\begin{equation}
\label{eq:Gab-T4-final}
{\cal G}_{a,b}^{(ec)}(V_b)=-\frac{e^2}{h} T^4 \frac{\xi}{l_e^{(S)}}
\frac{\Delta^2}{\Delta^2-(e V_b)^2}
\exp{\left[-\left(\frac{2 d}{\xi}\right)\right]}
,
\end{equation}
in agreement with
the expansion to order $T^4$ of the crossed
conductance obtained for highly transparent localized
interfaces\cite{Melin-Feinberg}.
A summation over the pairs of sites $\alpha$
and $\beta$ at the two
interfaces was carried out, giving rise to the
prefactor $\xi/l_e^{(S)}$ in Eq.~(\ref{eq:Gab-T4-final}).

\subsection{Weak localization diagram of order $T^6$}
\begin{figure}
\includegraphics [width=.8 \linewidth]{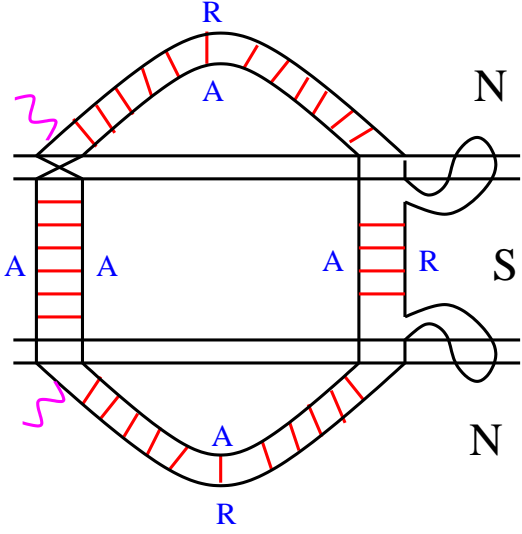}
\caption{(Color online.) 
Details of the insertion of the densities of states and of 
the $\hat{\sigma}_3$ matrices in the weak localization diagram on
Fig.~\ref{fig:A-bis}d. ``A'' and ``R'' stand for advanced and retarded.
The diffusons are calculated in the ladder approximation. The rungs of
the ladders represented by the red lines represent schematically
the diffuson in the ladder approximation. 
\label{fig:T6}
}
\end{figure}

\begin{figure*}
\includegraphics [width=.7 \linewidth]{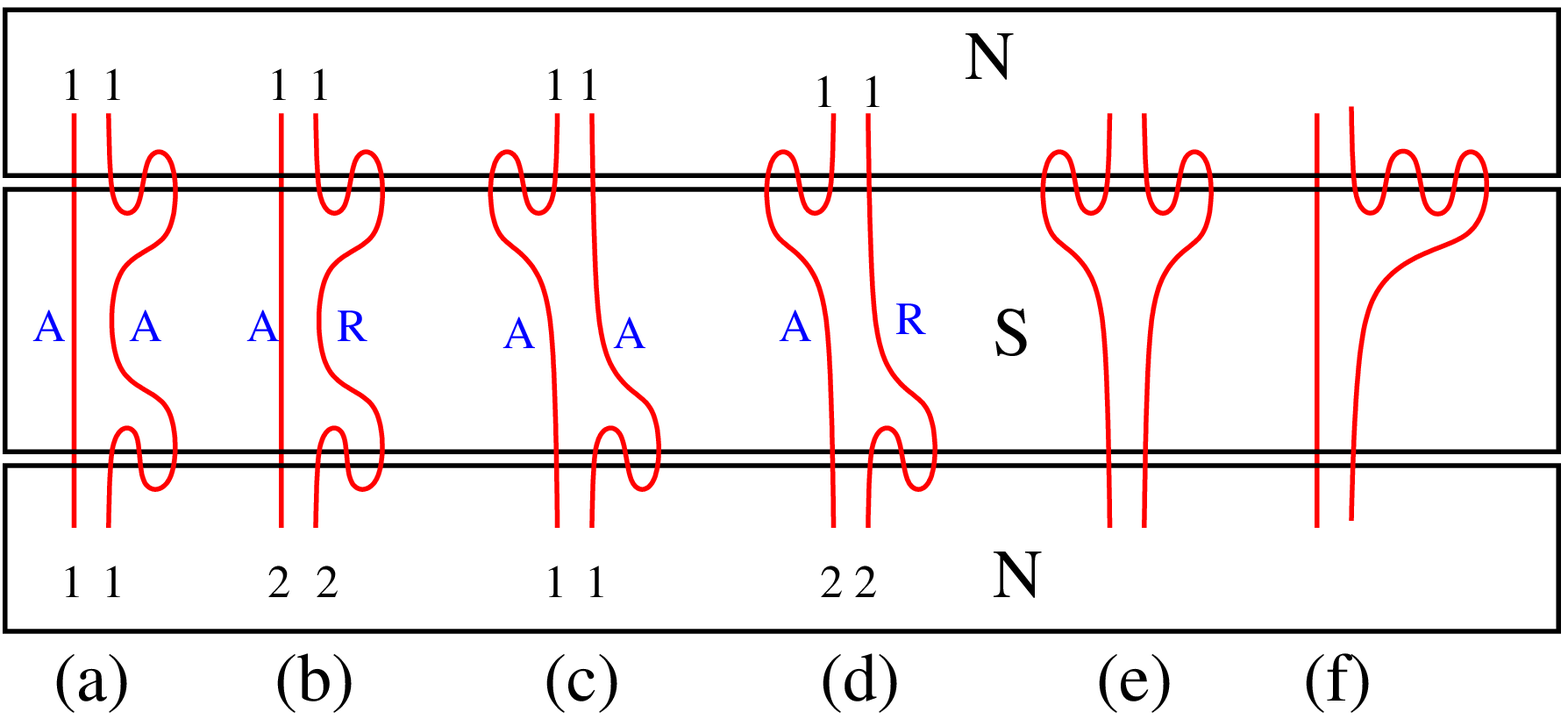}
\caption{(Color online.) Representation of the part of the diagram
of order $T^6$ (see Fig.~\ref{fig:A-bis})
with local excursions. ``1'' and ``2'' correspond to the Nambu labels.
(a), (b), (c) and (d) contribute respectively to
$X^{A,A}_{11,11}$, $X^{A,R}_{11,22}$,
$X^{A,A}_{11,11}$ and $X^{A,R}_{11,22}$.
The diagrams having the topology of (e) and (f) lead to a
vanishingly small crossed conductance because of the matrices
$\hat{\sigma}_3$ and the trace over the Nambu labels.
\label{fig:X}
}
\end{figure*}

\subsubsection{Contribution of local excursions}

We consider now the weak localization diagram of order $T^6$ on 
Fig.~\ref{fig:A-bis}d, merging the features of the two diagrams
of order $T^4$ on Figs.~\ref{fig:A-bis}b and c, and shown
in more details on Fig.~\ref{fig:T6}.
The long range propagation in the normal electrodes involves the
diffusons
$\overline{g^{A,1,1}_{\alpha,\beta} g^{R,1,1}_{\alpha,\beta}}$ 
and
$\overline{g^{A,2,2}_{\alpha,\beta} g^{R,2,2}_{\alpha,\beta}}$
in the particle-particle or hole-hole channel (where $\alpha$
and $\beta$ belong to the normal electrode),
as opposed to the diffuson
$\overline{g^{A,1,1}_{\alpha,\beta} g^{A,2,2}_{\alpha,\beta}}$
in the particle-hole channel
for local Andreev reflection at a single NIS interface
\cite{Hekking,Beenakker} below the related Thouless energy.

We first evaluate the part of the diagram on Fig.~\ref{fig:T6}
involving local excursions at the NIS interfaces (see Fig.~\ref{fig:X}).
Enumerating these diagrams shows that the two local
excursions are attached to the same non local propagation in
the superconductor.
We use the notation
$X^{A,A}_{\tau \tau,\tau' \tau'}$ ($X^{A,R}_{\tau \tau,\tau' \tau'}$)
for the part of the
diagram on Fig.~\ref{fig:X} with two advanced
(one advanced and one retarded) propagators in the superconductor,
and the Nambu labels $(\tau,\tau)$ and $(\tau',\tau')$
at the extremities in the two normal electrodes.
We find
\begin{eqnarray}
\label{eq:caca1}
X^{A,A}_{11,11} &=& X^{A,A}_{11,22}
= -\frac{8 \pi^2 \rho_N^6}{\mu(R)}
\frac{\omega^2 \Delta^2}{(\Delta^2-\omega^2)^2}\\
\label{eq:XAR11-11}
X^{A,R}_{11,11} &=& -X^{A,R}_{11,22}
= \frac{2 \pi^2 \rho_N^6}{\mu(R)}
\frac{\Delta^2}{\Delta^2-\omega^2}
.
\end{eqnarray}
We recover Eq.~(\ref{eq:Gab-T4-final}) for the non vanishing
diagram of order
$T^4$ on Fig.~\ref{fig:A-bis}c, proportional to
$-(X^{A,R}_{11,11}-X^{A,R}_{11,22})$. The global minus sign in
this expression can be found in Eq.~(\ref{eq:Gab-def}),
and the opposite signs for $X^{A,R}_{11,11}$
and $X^{A,R}_{11,22}$ are due to the matrices
$\hat{\sigma}_3$ in Eq.~(\ref{eq:Gab-def}).
The diagrams with local excursions attached to the disorder
average of two advanced or two retarded Green's functions
in the superconductor lead to a vanishingly small
crossed conductance, as it can be seen from the
identity $-(X^{A,A}_{11,11}-X^{A,A}_{11,22})=0$,
deduced from Eq.~(\ref{eq:caca1}).

\subsection{Weak localization crossed conductance}

The weak localization contribution
to the crossed conductance corresponding to the diagram on
Fig.~\ref{fig:T6} is given by
\begin{equation}
\label{eq:T4-T6}
\label{eq:G-wl}
{\cal G}_{a,b}^{(wl)}(e V_b) = - {\cal G}_{a,b}^{(ec)}(V_b)
F\left(
\frac{d}{\xi},
\frac{V_b}{\Delta},\frac{l_\varphi^{(N)}}{l_e^{(N)}}\right)
,
\end{equation}
where ${\cal G}_{a,b}^{(ec)}(V_b)$ is given by
Eq.~(\ref{eq:Gab-T4-final}) and
\begin{eqnarray}
\label{eq:y2D}
&&F\left(
\frac{d}{\xi},
\frac{V_b}{\Delta},\frac{l_\varphi^{(N)}}{l_e^{(N)}}\right) =\\
&&\frac{1}{4} T^2\left(\frac{l_\varphi^{(N)}}{l_e^{(N)}}\right)^4
\left(\frac{\xi}{l_e^{(S)}}\right)
\frac{\Delta^2}{\Delta^2-(eV_b)^2}
\exp{\left[-\left(\frac{2 d}{\xi}\right)\right]}
\nonumber
.
\end{eqnarray}
The sign of ${\cal G}_{a,b}^{(wl)}(e V_b)$ given by
Eq.~(\ref{eq:T4-T6}) is positive, as
for crossed Andreev reflection.
The factor $(l_\varphi^{(N)}/l_e^{(N)})^4$ is due to a a factor
$(l_\varphi^{(N)}/l_e^{(N)})^2$ associated to the diffusons in each
of the quasi-two-dimensional normal layers
[see Eq.~(\ref{eq:PNq})] in the limit $q=0$.
The constraint $q=0$ originates from the conservation of the component
of the wave-vector parallel to the interface, due to the symmetry
by translation parallel to the interfaces (see Appendix~\ref{app:B}).

It can be shown that the two diagrams of order $T^6$ involving a 
single diffuson in the superconductor and long range propagation
in the normal metals are negligible because of the sum over the
Nambu labels in one diagram, and because of the
factor $(l_\varphi^{(N)}/l_e^{(N)})^2$ in the other diagram, much
smaller than $(l_\varphi^{(N)}/l_e^{(N)})^4$ for the weak localization
diagram.

The weak localization crossed conductance can be expanded systematically
in powers of $(l_\varphi/l_e^{(N)})^2$:
\begin{equation}
\label{eq:expan}
{\cal G}_{a,b}(e V_b=0)=\frac{e^2}{h}
\sum_n A_n(T) \left(\frac{l_\varphi}
{l_e^{(N)}} \right)^{2 n}
.
\end{equation}
The coefficients $A_n(T)$ are evaluated to leading order in $T$
because of the small interface transparencies.
Estimating the
higher order weak localization diagrams
leads to $A_3 \sim -T^8$, $A_4=0$, $A_5 \sim T^{10}$, $A_6=0$,
$A_7 \sim -T^{12}$,~... 
The order of magnitude of
the sum of the higher order contributions can be obtained from
summing the corresponding
geometric series for $T (l_\varphi/l_e^{(N)})^2>1$:
$\sum_{n\ge 3} A_n(T) \left(l_\varphi/
l_e^{(N)} \right)^{2 n}$ is of order $T^2 \left(l_\varphi/
l_e^{(N)} \right)^{4}$, much smaller than
$A_2(T) \left(l_\varphi/
l_e^{(N)} \right)^{4}$, therefore justifying why we based our discussion
on the first two terms $A_0(T) \sim T^4$ and $A_2(T)\sim T^6$.

\subsection{Relation to experiments}
\label{subsec:exp}
\subsubsection{Determination of the parameters}
\label{sec:determin}
The number of channels
$N_{\rm ch}$ for a contact with a three-dimensional metal
is $N_{\rm ch}={\cal S}/\lambda_F^2$, where ${\cal S}$ is the
junction area and $\lambda_F$ the Fermi wave-length.
The normal layers have a dimension $L\times L \times d'$,
with $L\simeq 5\,\mu$m and a thickness $d'\simeq 50\,$nm
(see Fig.~\ref{fig:trilayer}).
The number
of channels in the quasi-two-dimensional geometry is
obtained be neglecting disorder (the elastic mean free
path in the experiment is limited by scattering on the normal
film boundaries), and by evaluating the area of the Fermi
surface with discrete wave-vectors in the direction
perpendicular to the interface, leading to
$N_{\rm ch}=L d'/\lambda_F^2\simeq 6\times
10^5$.
The normal transparency $T$
can be obtained from the local conductance in the normal state
${\cal R}_N^{({\rm loc})}\simeq 2\,\Omega$:
\begin{eqnarray}
\frac{1}{{\cal R}_N^{({\rm loc})}}\simeq 2 N_{\rm ch} T \frac{e^2}{h}
,
\end{eqnarray}
leading to $T\simeq 10^{-2}$. 
These values $T\simeq 10^{-2}$ and $N_{\rm ch}\simeq 6\times 10^5$ are
compatible with the local Andreev resistance at
zero bias of about $100\,\Omega$, being an upper bound to the
zero temperature Andreev resistance ${\cal R}_A^{({\rm loc})}$
given by $1/{\cal R}_A^{({\rm loc})}=2(e^2/h) T^2 N_{\rm ch}$.
Russo {\it et al.} \cite{Russo}
estimate $T\simeq 10^{-5}$
from $N_{\rm ch}={\cal S}/\lambda_F^2$ for a three dimensional metal.
The possible
dependence of
$N_{\rm ch}$ on $d'$ that we consider here
can be probed experimentally by determining how the
crossed resistance
depends on the thickness of the normal layers.

\subsubsection{Crossed resistance}
\label{sec:crossed-resistance}
The crossed resistance matrix measured experimentally \cite{Russo}
is the inverse of the crossed conductance matrix. The off-diagonal
element of the crossed resistance matrix is given by
\begin{equation}
\label{eq:crossed-R}
{\cal R}_{a,b}^{({\rm tot})}(V_b)=-
\frac{{\cal G}_{a,b}^{({\rm tot})}(V_b)}
{{\cal G}_{a,a}^{({\rm tot})}(V_b)
{\cal G}_{b,b}^{({\rm tot})}(V_b)-{\cal G}_{a,b}^{({\rm tot})}(V_b)
{\cal G}_{b,a}^{({\rm tot})}(V_b)}
,
\end{equation}
where 
${\cal G}_{a,a}^{({\rm tot})}(V_b)=
{\cal G}_{b,b}^{({\rm tot})}(V_b)={\cal G}_{\rm loc}^{({\rm tot})}(V_b)$
is the local Andreev conductance.

The elastic cotunneling
crossed resistance corresponding solely to the contribution
of Eq.~(\ref{eq:Gab-T4-final}) is thus of order
\begin{eqnarray}
\nonumber
{\cal R}_{a,b}^{({\rm ec})}(V_b=0)
&=&
\frac{1}{4 N_{\rm ch}} \frac{h}{e^2} 
\left(\frac{\xi}{l_e^{(S)}}\right)\\
&\times&
\left(\frac{\Delta^2-(e V_b)^2}{\Delta^2}\right)
\exp{(- 2 d/\xi)}
,
\end{eqnarray}
having an order of magnitude compatible
with the experiment \cite{Russo}. 
The elastic cotunneling crossed resistance is independent on
$T$ for tunnel interfaces.
The elastic cotunneling crossed resistance
${\cal R}_{a,b}^{({\rm ec})}(V_b=0)$ being inversely proportional
to $N_{\rm ch}$, with $N_{\rm ch}$ proportional to $d'$
(see Sec.~\ref{sec:determin}), it is expected that the crossed
resistance would decrease if the normal metal layer thickness $d'$
increases, as anticipated in Sec.~\ref{sec:determin}.

Now, the weak localization crossed resistance is equal to
\begin{equation}
{\cal R}_{a,b}^{(wl)}=-{\cal R}_{a,b}^{(ec)}
F\left(\frac{d}{\xi},
\frac{V_b}{\Delta},\frac{l_\varphi^{(N)}}{l_e^{(N)}}\right)
.
\end{equation}
\begin{figure}
\includegraphics [width=1. \linewidth]{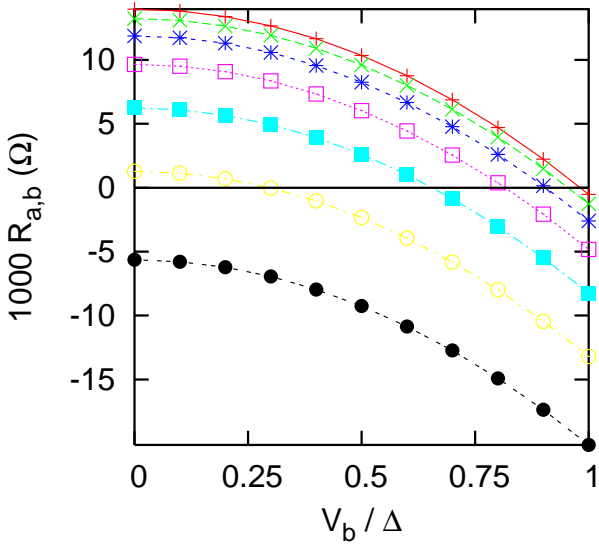}
\caption{(Color online.) Voltage dependence of the crossed resistance
(in Ohms),
given by Eq.~(\ref{eq:crossed-R}), in which the crossed conductance is
the sum of the elastic cotunneling [see Eq.~(\ref{eq:Gab-T4-final})]
and weak localization [see Eq.~(\ref{eq:G-wl})] contributions.
We used the parameters $d=20\,$nm, $\xi=15\,$nm,
$l_e^{(N)}=50\,$nm, $l_e^{(S)}=3\,$nm, $T=10^{-2}$,
$N_{\rm ch}=6 \times 10^5$. The values of $l_\varphi^{(N)}$ are, from
top to bottom:
$l_\varphi^{(N)}=0.4,\,0.5,\,0.6,\,0.7,\,0.8,\,0.9,\,1\,\mu$m.
The points are theoretical, and the lines are a guide to the eye.
\label{fig:R_vs_V}
}
\end{figure}
The voltage dependence of the total crossed resistance 
${\cal R}_{a,b}(V_b)={\cal R}_{a,b}^{(ec)}(V_b)
+{\cal R}_{a,b}^{(wl)}(V_b)$
is shown
on Fig.~\ref{fig:R_vs_V} for different values of
$l_\varphi^{(N)}/l_e^{(N)}$. We obtain a change of sign from
a positive (elastic cotunneling dominated) to a negative
(weak localization dominated) crossed resistance as the voltage increases
(see Fig.~\ref{fig:R_vs_V}).
For sufficiently large values of the phase coherence length, we have
$F\left(d/\xi,0,l_\varphi^{(N)}/l_e^{(N)}\right)>1$ as discussed above,
so that the crossed resistance is negative for all values of the
bias voltage. 

The perturbative crossed resistance on Fig.~\ref{fig:R_vs_V}
tends to a finite value in the limit $e V_b=\Delta$. The determination
of the crossed conductance around $e V_b=\Delta$ is examined in
Ref.~\onlinecite{Melin-Feinberg} for localized interfaces
with arbitrary interface transparencies.
The non perturbative crossed conductance tends to
zero for $e V_b=\Delta$, but the cross-over occurs
within an energy window that
becomes very small for tunnel interfaces. A crossed current
related to out-of-equilibrium quasiparticle populations
(not described here)
is expected for $e V_b>\Delta$. 

As it is visible on Fig.~\ref{fig:R_vs_V}, the characteristic
voltage scale in the bias voltage dependence of the crossed resistance
is the superconducting gap,
not the normal state
superconductor Thouless energy obtained in experiments. At the present
stage, we do not find a plausible explanation of this experimental
observation.

\subsection{Magnetic field dependence}
The experimental crossed signal is suppressed by a magnetic 
field parallel to the layers \cite{Russo}. 
The theoretical weak localization crossed signal is also suppressed by
a magnetic field because the corresponding diagram
dephases in an applied magnetic field.
The cross-over magnetic field $B_*$
for the suppression of the weak localization crossed conductance
corresponds to one
superconducting flux quantum $\phi_0$ enclosed
in the area of the diagram, compatible with experiments \cite{Russo}.
However, in experiments, the crossed resistance is suppressed
by a magnetic field in the entire voltage range.
The present model does not explain the dephasing of the elastic
cotunneling contribution. 

\section{Conclusions}
\label{sec:conclu}
We have calculated the weak localization contribution to non local
transport
in NISIN trilayers with extended interfaces and
a sufficient phase coherence length in the normal electrodes.
We find a change of sign in the crossed resistance
between elastic cotunneling at low
voltages and weak localization at higher voltages. The weak localization
contribution can dominate for all voltages if the phase coherence length
is large enough. The weak localization crossed conductance dephases in
an applied magnetic field, but not the elastic cotunneling contribution.
The appearance of a voltage scale related to the superconductor
Thouless energy is left as an important open question.

The author thanks H. Courtois, A. Morpurgo,
and S. Russo
for many useful suggestions and comments.
Special thanks to D. Feinberg and F. Pistolesi for numerous
fruitful discussions.

\appendix

\section{Diffusion probabilities in a normal metal}
\label{sec:diff-prob}
\subsection{Diffusion in the electron-electron channel}
We first discuss briefly the normal metal diffuson in the ladder
approximation. In the Born approximation,
the elastic scattering time $\tau_{1,1}$
in the ``11'' Green's function given by
\begin{equation}
g_{1,1}^{A,R}({\bf k},\omega)=
\frac{1}{\omega-\xi_k\mp i/\tau_{1,1}}
,
\end{equation}
is defined by
$k_F^3 \tau_{1,1} v^2/(4\pi\epsilon_F)=1$,
with $\tau_{1,1}=\tau_0 [1+\omega/(2\epsilon_F)]$,
where $\tau_0$ is the elastic scattering time at the Fermi level,
$v$ is the amplitude of the microscopic impurity scattering
potential, $k_F$ is the Fermi wave-vector, and 
$\epsilon_F$ the Fermi energy.

Using contour integration, and the identity
\begin{equation}
\int \frac{d {\bf k}}{(2\pi)^3} f({\bf k})
=\frac{1}{8\pi^2} \int_{-\infty}^{+\infty}
k^2 dk \int_{-1}^{1} du  f(k,u) e^{i k R u}
,
\end{equation}
with $k=\pm |{\bf k}|$ and $f({\bf k})$ a function of
${\bf k}$, we obtain easily
\begin{eqnarray}
\label{eq:identite-int}
&&\int \frac{d {\bf k}}{(2\pi)^3}
g_N^{1,1}({\bf k}) g_N^{1,1}({\bf k}+{\bf q})\\
&=&\frac{k_F^3 \tau_{1,1}}{4\pi\epsilon_F}
\left(1-\frac{D \tau_0}{4}\left[q^2+(l_\varphi^{(N)})^{-2} \right]
\right)
\nonumber
,
\end{eqnarray}
where $D$ is the diffusion constant.
The Fourier transform ${\cal P}^N(q)$ of the diffusion probability
${\cal P}^N({\bf r})$, given by
\begin{equation}
{\cal P}^N(q)=t_0^2 \int \frac{d {\bf k}}{(2\pi/a_0)^3}
\overline{
G^A_{1,1}({\bf k},\omega) G^R_{1,1}({\bf k}+{\bf q},\omega)
}
,
\end{equation}
where the overline is a disorder averaging and $t_0$ is the bulk
hopping integral (see Eq.~(\ref{eq:HN})),
is obtained by summing the ladder series, leading to
\begin{equation}
\label{eq:PNq}
{\cal P}^N({\bf q})=\frac{4}{D \tau_0} \frac{1}
{{\bf q}^2+(l_\varphi^{(N)})^{-2}}
.
\end{equation}
In this expression ${\cal P}^N(q)$ has no dimension.
Its Fourier transform ${\cal P}^N({\bf R})$ is
such that ${\cal P}^N({\bf R}) d {\bf R}$ has
no dimension, as expected for a probability. 

\section{Double NS interface}
\label{app:B}

We provide now explanations to
the factor $(l_\varphi^{(N)}/l_e^{(N)})^4$
appearing in Eq.~(\ref{eq:y2D}) for quasi-two-dimensional
normal electrodes. 
Because of the
conservation of the component of the wave-vector parallel to the
interface at each tunneling vertex, and because of the form
of the diagram, the diagram on Fig.~\ref{fig:T6}
is evaluated at ${\bf q}_\parallel=0$, where ${\bf q}_\parallel$
is the projection of ${\bf q}$ (see Appendix~\ref{sec:diff-prob})
on a plane parallel to the interfaces.
By momentum conservation, a finite value of ${\bf q}_\parallel$ is
transformed in $-{\bf q}_\parallel$ after traversing the entire diagram,
so that ${\bf q}_\parallel=-{\bf q}_\parallel=0$.
The resulting crossed conductance is thus
proportional to $\left[{\cal P}^N({\bf q}_\parallel=0)\right]^2$, where
the square is due to the correlated diffusion in the two normal
electrodes, and where the diffusion probability
${\cal P}^N({\bf q}_\parallel)$ is given by
Eq.~(\ref{eq:PNq}). This lead to the factor
$\sim (l_\varphi^{(N)}/l_e^{(N)})^4$ since the normal metals
are quasi-two-dimensional so that $q_\perp=0$, where $q_\perp$
is the projection of ${\bf q}$ on the normal to the interfaces.

\section{Ballistic NISIN trilayer with atomic thickness}
We consider in this Appendix a ballistic NISIN trilayer
with atomic thickness \cite{Buzdin-ato,Melin-Feinberg-ato}
in which the three electrodes are two-dimensional, and
find a factor $(l_\varphi^{(N)}/l_e^{(N)})^3$, similar
to the factor
$(l_\varphi^{(N)}/l_e^{(N)})^4$ factor discussed 
previously in the diffusive limit.

The transport 
formula given by Eq.~(\ref{eq:Gab-def}) is Fourier transformed,
to obtain
\begin{eqnarray}
\label{eq:Gab-k}
&&{\cal G}_{a,b}(V_b) = -8 \pi^2 \frac{e^2}{h} t^4 
\int \frac{d^2{\bf k}}{(2\pi)^2} \\
&& \mbox{Tr}
\left\{ \hat{\rho}^N_{a,a}({\bf k})
 \hat{\sigma}_3
\hat{G}_{\alpha,\beta}^{A}({\bf k},e V_b)
\hat{\rho}^N_{b,b}
\hat{\sigma}_3
\hat{G}_{\beta,\alpha}^{R}({\bf k},e V_b)
\right\}
,
\nonumber
\end{eqnarray}
and the Dyson equations for
$\hat{G}^{A,R}({\bf k})$ are also Fourier transformed.
The normal metal ballistic Green's function is given by
$g_{a,a}^{1,1}({\bf k},\omega)=1/[\omega-\xi+i \eta]$,
where $\xi$ is the kinetic energy with respect to the
Fermi level, and where the broadening parameter $\eta$ is
given by $\eta=\hbar v_F l_\varphi^{(N)}$.
Eq.~(\ref{eq:Gab-k}) is then expanded diagrammatically.
A diagram similar to the one
on Fig.~\ref{fig:T6} leads to a factor $\eta^{-3}
= (\hbar v_F l_\varphi^{(N)})^3$.


\begin{thebibliography}{99}
\bibitem{note-R} The crossed resistance ${\cal R}_{a,b}(V_b) =
  \partial V_a/\partial I_b(V_b)$, measured experimentally
  in Ref.~\cite{Russo}, has a sign opposite to the crossed conductance.
\end{thebibliography}
\end{document}